\documentclass[prb,twocolumn,showpacs,superscriptaddress]{revtex4}

\usepackage{graphicx}
\usepackage{amsmath}
\usepackage{amssymb}
\usepackage{epsfig}

\renewcommand{\v}[1]{{\bf #1}}

\def\eqa{\begin{eqnarray}}
\def\eea{\end{eqnarray}}
\newcommand{\eq}{\begin{equation}}
\newcommand{\ee}{\end{equation}}
\newcommand{\nn}{\nonumber\\}

\newcommand{\<}{\langle}
\renewcommand{\>}{\rangle}

\newcommand{\p}{\partial}

\newcommand{\ua}{\uparrow}
\newcommand{\da}{\downarrow}

\newcommand{\al}{\alpha}
\newcommand{\bt}{\beta}

\newcommand{\Del}{\Delta}

\newcommand{\ga}{\gamma}
\newcommand{\Ga}{\Gamma}

\newcommand{\la}{\lambda}
\newcommand{\La}{\Lambda}

\newcommand{\si}{\sigma}

\usepackage{color}

\begin{document}

\title{Antiferromagnetism, $f$-wave and chiral $p$-wave superconductivity in a Kagome lattice with possible application to $sd^2$-graphenes}

\author{Wan-Sheng Wang}
\affiliation{National Laboratory of Solid State Microstructures $\&$ School of Physics, Nanjing
University, Nanjing, 210093, China}
\affiliation{Department of Physics, Ningbo University, Ningbo 315211, China}

\author{Yuan-Chun Liu}
\affiliation{National Laboratory of Solid State Microstructures $\&$ School of Physics, Nanjing
University, Nanjing, 210093, China}

\author{Yuan-Yuan Xiang}
\affiliation{College of Science, Hehai University, Nanjing, 210098, China}

\author{Qiang-Hua Wang}
\affiliation{National Laboratory of Solid State Microstructures $\&$ School of Physics, Nanjing
University, Nanjing, 210093, China}
\affiliation{Collaborative Innovation Center of Advanced Microstructures, Nanjing 210093, China}
\email{qhwang@nju.edu.cn}

\begin{abstract}

We investigate the electronic instabilities in a Kagome lattice with Rashba spin-orbital coupling by the unbiased singular-mode functional renormalization group. At the parent $1/3$-filling, the normal state is a quantum spin Hall system. Since the bottom of the conduction band is near the van Hove singularity, the electron-doped system is highly susceptible to competing orders upon electron interactions. The topological nature of the parent system enriches the complexity and novelty of such orders. We find $120^o$-type intra-unitcell antiferromagnetic order, $f$-wave superconductivity and chiral $p$-wave superconductivity with increasing electron doping above the van Hove point. In both types of superconducting phases, there is a mixture of comparable spin singlet and triplet components because of the Rashba coupling. The chiral $p$-wave superconducting state is characterized by a Chern number $Z=1$, supporting a branch of Weyl fermion states on each edge. The model bares close relevance to the so-called $sd^2$-graphenes proposed recently.

\end{abstract}

%

\maketitle

\section{Introduction}

The Kagome lattice systems have attracted considerable attentions due to the rich physics
associated with the high degree of geometrical frustration. In the Mott insulating limit, several
possible states have been proposed as the ground states of the Heisenberg model
in this lattice, such as a valance bond solid state (with $36$ sites per unit cell),\cite{ZengC, Huse, Balents, Vidal, Poilblanc}
a U(1) algebraic spin liquid (SL),\cite{WenXG}
a disordered triplet gapped SL,\cite{WengZY} and a singlet gapped SL with
signatures of $Z_2$ topological order and anyonic excitations.\cite{JiangHC, White, Depenbrock}
At low hole doping, a special valance bond crystal state (with
12 sites per unit cell) is proposed as the ground state of the $t$-$J$ model in
this lattice. \cite{Guertler1,Guertler2} For intermediate interactions and at
various fillings, several exotic phases have been proposed for the Kagome
lattice, such as peculiar Mott transition, \cite{Ohashi, Lauchli}
anomalous quantum Hall effects,\cite{Ohgushi, Petrescu} and semionic states
at half filling,\cite{Bauer} fractional charge at $1/3$ filling for
spinless fermions,\cite{Brien} and ferromagnetism at electron filling
$1/3$ (or $5/3$).\cite{Pollmann}

There are also extensive interests in the itinerant limit of the Kagome lattice.
In particular, at the van Hove filling, the Fermi surface (FS)
is perfectly nested and has saddle points on the edges of the Brillouine Zone (BZ).
This would make the system unstable against weak interactions. Similar FS appears in
triangle and honeycomb lattices and was shown to develop a chiral spin-density-wave (SDW)
,\cite{Martin, LiTao, wws1} or a chiral $d_{x^2-y^2} + i d_{xy}$ pairing.\cite{Chubukov,Thomale1}
Both states break time-reversal and parity symmetries and are topologically nontrivial.
Given the similar FS, a simple FS nesting argument would predict similar orders
in the Kagome lattice. However, the character of the Bloch state on the FS depends on the
position of the momentum. The nesting vector connects two Bloch states with different sublattice contents,
weakening the nesting effect as far as an on-site interaction $U$ is concerned.\cite{LJX,wws2,Thomale2}
Such a matrix element effect causes profound difference to the case of honeycomb lattice.
At the van Hove filling, the instabilities are ferromagnetism, intra-unitcell
antiferromagnetism and charge-bond-order (CBO) with increasing $U$.\cite{wws2}
There is a concern on the fate of the ferromagnetic order in the limit of large $U$.\cite{Thomale3}
However, such an order is unlikely as viewed from the equivalent $t$-$J$ model, which supports the CBO
instead.\cite{Guertler2}
In addition, the nearest-neighbor (NN) repulsion $V$ connects different sublattices and thus enjoys the FS nesting,
leading to additional instabilities. For example, the $d_{x^2-y^2} + i d_{xy}$ pairing and spin-bond-order
appear if $V$ and $U$ are comparable, while the charge-density-wave (CDW) and $s$-wave pairing are
found if $V$ dominates.\cite{wws2}

The orders mentioned so far are found in the absence of spin-orbital coupling (SOC). We ask how would SOC
modify or enrich the orders. We are motivated by the fact that SOC could cause nontrivial topology already
at the single-particle level. In the Kane-Mele model, the Rashba-type SOC connecting next-nearest-neighbor (NNN)
bonds of the honeycomb lattice, and makes the system at half filling a quantum spin Hall (or a two-dimensional
topological insulator). However, in a realistic system such as graphene, the SOC is too small to be practical.
Apart from the lightness of the carbon atom, the weakness of SOC in graphene arises from the fact that the SOC
connects the longer NNN bond about which the two sides are asymmetric. In Kagome lattice, however, there is an asymmetry
already about the NN bond. It is therefore more hopeful to look for SOC in Kagome lattices.
Interestingly, such a material is indeed proposed from first principle. It is a honeycomb lattice of transition metal
(TM) atoms, but the low energy electrons are described effectively by Kagome bands.\cite{MZhou} This is because the
$sd^2$-hybridized $\si$-orbitals are bond-centered on the honeycomb lattice, so that electrons hop effectively on
the Kagome lattice dual to the honeycomb lattice. The TM atoms display large Rashba-type SOC. Such a system is dubbed
$sd^2$-graphene. The system displays magnetism (from local moment due to unsaturated lower atomic levels) for the TM
element W, enabling quantum anomalous Hall effect.\cite{MZhou} The calculation also shows that the atomic magnetism
could be avoided for other TM elements. Here we are interested in the latter case, as we are concerned mainly on the
superconducting order, which would be unlikely stable against the strong local spin exchange.
We notice that in the parent $sd^2$-graphene (at the $1/3$ filling), the bottom of the conduction band above the band gap is rather flat
and close to the van Hove point. Therefore by slight electron doping the system would be highly susceptible to competing orders
upon electron interactions. The topological nature of the parent system would enrich the complexity and novelty of such orders.

In this paper, we investigate the electronic instabilities of electron doped $sd^2$-graphene near the van Hove filling.
To treat the correlation effects in an unbiased way, we resort to the singular-mode functional renormalization group (SMFRG)
developed recently.\cite{wws1,xyy1,wws3} We find intra-unitcell antiferromagnetism (AFM), $f$-wave and chiral
$p+ip$-wave superconductivity (SC), with increasing deviation from the van Hove level. In both types of superconducting phases, there is a mixture of
comparable spin singlet and triplet components because of the Rashba coupling. The chiral $p$-wave superconducting
state is characterized by a Chern number $Z=1$, supporting a branch of Weyl fermion states on each edge.

The rest of the paper is arranged as follows. In Sec.\ref{MM}, we define the model and
briefly introduce the SMFRG method. In Sec.\ref{RD}, we discuss the results for various electron doping levels and present
the phase diagram. Finally, Sec.\ref{SP} is a summary and perspective of this work.

 \section{Model and Method} \label{MM}

 \begin{figure}
\includegraphics[width=8.5cm]{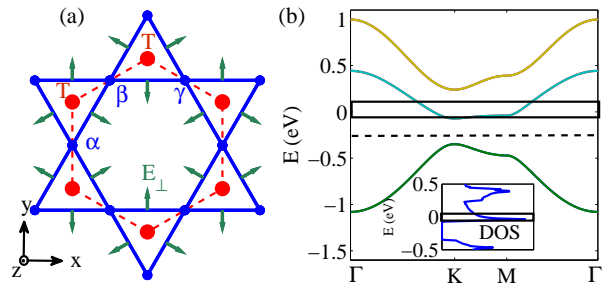}
\caption{ (Color online) (a) The structure of $sd^2$-graphene. The TM atoms (red circles) are located on the honeycomb lattice (red dashed lines).
The $\si$ orbitals (blue circles) define a Kagome lattice (blue lines), with three sublattices $\al$, $\bt$
 and $\ga$. The arrows show the crystal electric field $E_\perp$ on the NN bonds of the Kagome lattice.
 The spatial axes are defined by $x$, $y$ and $z$. (b) The
 band structure of the effective model. The dashed line indicates the Fermi level for the parent $1/3$ filling. The
 black rectangle shows the electron doping regime under concern. There is a van Hove singularity with large density of states near the
 bottom of the conduction band (inset).} \label{model}
\end{figure}

The structure of $sd^2$-graphene is schematically shown in Fig.\ref{model}(a),
where the red cycles and red-dashed lines show the position of TM atoms and the
corresponding honeycomb lattice. For TM atoms, the $s$, $d_{xy}$ and $d_{x^2-y^2}$ orbitals hybridize to form
bonding $\si$-orbitals around the centers of the NN bonds on the honeycomb lattice,
shown as solid blue cycles in Fig.\ref{model}(a),
as well as antibonding $\si^*$-orbitals at higher energy beyond the present interest.
In contrast to the $\pi$-orbitals in graphene, the bond-centered $\si$-orbitals form
effectively a Kagome lattice (dual to the honeycomb lattice), shown as the blue solid lines in Fig.\ref{model}(a), with three sublattices
$\al$, $\bt$, and $\ga$. The effective Hamiltonian derived from the $\si$-orbitals is given by \cite{MZhou}

\eqa  H = &-& \sum_{\< ij \> \si}(c^\dag_{i \si}t_{ij} c_{j \si}+{\rm h.c.})-\mu\sum_{i\si}n_{i\si}+U\sum_i n_{i\ua}n_{i\da}
 \nn
        &+& \frac{i\la}{2} \sum_{\< ij \>\in {\rm NN}, \si }( c_{i\si}^\dag \si \eta_{ij}  c_{j\si}-{\rm h.c.}).  \label{H}
\eea
Here $c^\dag_{i\si}$ creates an electron with spin $\si=\pm 1$, $\mu$ is the chemical potential, and $\<ij\>$ denotes the bonds
between the first, second and third neighbors (henceforth on Kagome lattice), with the corresponding hopping integrals $t_{1,2,3}$.
The Rashba SOC on the NN bonds is given by $\la$ in strength, and $\eta_{ij}=\pm 1$ is the sign of $\hat{z}\cdot(\v E_{ij}\times\v R_{ij})$,
where $\v E_{ij}$ is the crystal electric field, depicted as arrows in Fig.\ref{model}(a), and $\v R_{ij}=\v r_i-\v r_j$ is
the displacement vector during electron hopping.

For concreteness, we set $t_1 = 0.25$eV, $t_2 = 0.05$eV, and $t_3 = -0.02$eV
and we set Rashba SOC $\la = 0.16$eV, as it is found strong in Ref.\cite{MZhou}. (Our results do not change qualitatively against fine tuning
of these parameters, as will be discussed in the end of section III.)
The corresponding band structure is shown in Fig.\ref{model}(b). Notice that each band is two-fold degenerate even in the presence of the SOC
in Eq.\ref{model}. At the parent $1/3$ filling, the Fermi level lies between the lower two bands.
We are interested in electron doping above the van Hove singularity in the conduction band.
We should emphasize that the hopping integrals may depend on the TM atoms, but the existence of a van Hove point near the bottom of the
conduction band is quite general. On the other hand, the doping level can be controlled by TM substitution,
adsorption of alkaline atoms, or electric gating.

We treat the effect of the Coulomb interaction $U$ by SMFRG. \cite{wws1, xyy1, wws3}
A general four-point interaction vertex $\Ga$ can be decomposed as
\eq  \Ga = \sum_{\v q \nu} (A^\nu_{\v q})^\dag S^\nu(\v q) A^\nu_{\v q}, \ee
where $\nu$ denotes an eigenmode of $\Ga$, $\v q$ is a collective momentum,
and $A^\nu_{\v q}$ is a composite boson field made up of a fermion bilinear. In the particle-particle (pp)
channel, $(A^\nu_{\v q})^\dag = \sum_{\v k} \psi^\dag_{\v k + \v q}
\phi^\nu_{\v q}(\v k) (\psi^\dag_{-\v k})^T$,
while in the particle-hole (ph) channel, $(A^\nu_{\v q})^\dag =
\sum_{\v k} \psi^\dag_{\v k + \v q} \phi^\nu_{\v q}(\v k) \psi_{\v k}$.
Here $\psi_\v k$ is a six-component spinor in the
momentum space, made of $c_{\v k\si a}$ for $\si=\pm$ and $a=\al$, $\bt$ and $\ga$. The inner structure of
the composite boson is described by a matrix form factor $\phi^\nu_\v q(\v k)$.
The decompositions of the same interaction vertex into pp and ph channels imply that there
are overlaps between them, which are handled on equal footing in the SMFRG. We also emphasize that SMFRG
takes all bands into account, since it works in the orbital rather than band basis.
Stating form the bare interaction vertex given by the $U$-term in Eq.\ref{H}, the effective vertex $\Ga$
flows versus a decreasing infrared cutoff energy scale $\La$. We monitor the associated evolution of $S^\nu_{pp, ph}(\v q)$
as well as the form factors. The most negative (i.e., attractive) $S^\nu_{pp, ph}(\v q)$
indicates the leading mode in the respective channel, which we denote as $S_{pp, ph}$ for brevity.
A diverging leading mode implies an instability of the normal state, and the associated collective momentum $\v Q$ and form
factor $\phi(\v k)$ (dropping the collective momentum and the mode index) describe the emerging order parameter. The divergence energy scale is
an upper limit of the transition temperature $T_c$. More technical details can be found elsewhere. \cite{wws1,xyy1, wws3,wd}

\section{Results and Discussions} \label{RD}

In this section, we provide the SMFRG results for the model defined
in the previous section. We begin by discussing the results for specific
fillings (chemical potentials) with $U = 1$eV, and summarize the results by a phase
diagram in the $(U, \mu)$ space.

\begin{figure}
\includegraphics[width=8.5cm]{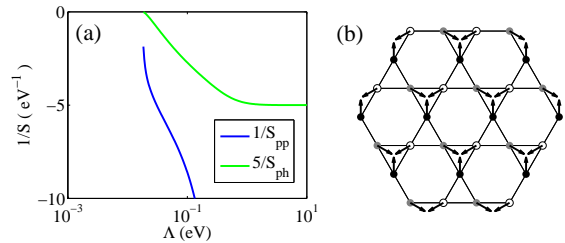}
\caption{ (Color online) Results for $\mu = -0.04$eV. (a) FRG flow of
(the inverse of) the most negative singular values $S$ in the particle-particle (pp)
and particle-hole (ph) channels. (b) the spin structure of the leading mode in
ph channel.}\label{AFM}
\end{figure}

{\em Intra-unit-cell antiferromagnetism}: We first consider the case of van Hove filling at which
$\mu = -0.04$eV. The FRG flow versus the running energy scale $\La$ is shown in
Fig.\ref{AFM}(a). Clearly, the ph channel is leading. We find that the associated collective momentum is
$\v Q = 0$, and the form factor $\phi(\v k)$ is given by, up to a global SO(2) symmetry within the $xy$-plane,
\eqa  (\phi^{\al \al},\phi^{\bt\bt},\phi^{\ga\ga}) =( -\si_y ,  \frac{\si_y + \sqrt{3} \si_x}{2}, \frac{\si_y - \sqrt{3} \si_x}{2}), \eea
where $\si_{x,y,z}$ are Pauli matrices in the spin basis. (The other elements are essentially zero.)
The form factor is diagonal in sublattice basis and independent of $\v k$, meaning the order is site-local. On the other hand,
the spin dependence indicates that the order is in the spin sector, and moreover the angle between nearby spins is $120^o$.
Finally, $\v Q=0$ means the spin ordering is ferromagnetic in terms of translation by unit cells.
The corresponding spin structure is shown in Fig.\ref{AFM}(b),
which we call an intra-unitcell AFM. This $\v Q=0$ instability in the ph channel is clearly enhanced by the van Hove singularity
rather than FS nesting, as a result of the matrix element effect discussed previously. It also appears even if the SOC is
absent.\cite{wws2}

\begin{figure}
\includegraphics[width=7.5cm]{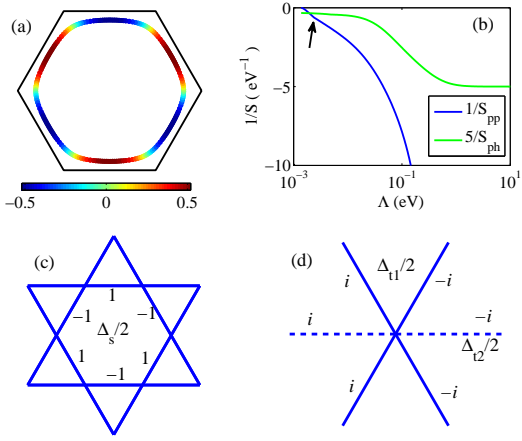}
\caption{ (Color online) Results for $\mu=-0.005$eV. (a) The Fermi surface and the gap
function (color scale) thereon. (b) FRG flow of (the inverse of) the most
negative singular values $S$ in the pp and ph channels.
The arrow indicates a level crossing of the paring modes.
(c) Real-space pattern of singlet pairing on NN bonds. The phase factor is put on bonds and the amplitude is $\Del_s/2$.
(d) Real-space pattern of triplet pairing on third-neighbor bonds, starting from an $\al$-site at the center.
The amplitudes on solid (dashed) bonds are $\Del_{t1}/2$ and $\Del_{t2}/2$, respectively. The difference arises from the
fact that the solid bond traverses a lattice site at the midpoint while the dashed one does not.
}\label{f}
\end{figure}

{\em $f$-wave superconductivity}: Now we consider a slightly higher electron doping with
$\mu = -0.005$eV. The FS is shown in Fig.\ref{f}(a), and Fig.\ref{f}(b) shows the FRG flow.
In the present case, the ph channel is enhanced at higher energy scales, but saturates at low energy scales
due to lack of phase space for low energy particle-hole excitations in the absence of perfect nesting.
On the other hand, the pp channel is triggered attractive (even though the initial interaction is repulsive) while the ph channel is enhanced,
a manifestation of the channel overlap. At still lower energy scales, the pp channel (at $\v Q=0$) can diverge on its own due to the
Cooper instability mechanism. Therefore, the system will develop superconductivity below
the divergence energy scale. The form factor in this case is just the pairing function,
which can be written as $\phi(\v k) = ( g_{\v k} +  d_\v k\si_z) i \si_y$ in our case,
with a singlet part $g_{\v k}$ and a triplet part $d_{\v k}$. (The `d-vector' is always oriented along $z$.) The non-zero
elements of $g_{\v k}$ and $d_\v k$ are given by
\eqa
  g^{\al \bt}_{\v k} &=& - g^{\bt \al}_{\v k} =  -i\Del_s \sin \frac{k_+}{2}, \nn
  g^{\al \ga}_{\v k} &=& - g^{\ga \al}_{\v k} =   i\Del_s \sin \frac{k_-}{2}, \nn
  g^{\bt \ga}_{\v k} &=& - g^{\ga \bt}_{\v k} =   i\Del_s \sin \frac{k_x}{2},\nn
  d^{\al \al}_{\v k} &=&  \Del_{t1}(\sin k_+ + \sin k_- ) +  \Del_{t2} \sin k_x, \nn
  d^{\bt \bt}_{\v k} &=&  \Del_{t1}(\sin k_+ - \sin k_x ) -  \Del_{t2} \sin k_-, \nn
  d^{\ga \ga}_{\v k} &=&  \Del_{t1}(\sin k_- - \sin k_x) -  \Del_{t2} \sin k_+,
\eea
where $k_{\pm} = (k_x \pm \sqrt{3} k_y)/2$. Notice that the singlet/tripet part corresponds to pairing on the first/third neighbor bonds.
We find that due to the SOC the singlet and
triplet components are mixed with comparable amplitudes $\Del_s : \Del_{t1} : \Del_{t2} \sim 7 : 8 : 6$. Such a pairing configuration enjoys simultaneously the quasi-AFM
correlation between nearby sublattices and FM correlation between like-sublattices (connected by the third-neighbor bonds), as discussed previously. The real-space pattern for the singlet and triplet parts are illustrated in Fig.\ref{f}(c) and (d), respectively. By inspection, the singlet part transforms as $f$-wave with respect to the symmetry center (the center of the holo hexagon), which would be invalid without the
sublattice structure. The same symmetry is carried by $d_\v k$, guaranteed by group theory, but less obvious in Fig.\ref{f}(d). To see this better,
we transform the pairing function into the band basis, $\Del_\v k = \< \v k | \phi(\v k) | \bar{\v k} \>$, where $| \v k \>$ is a Bloch state (in the normal state)
and $ | \bar{ \v k} \> = (T | \v k \>)^*$, with $T$ the time-reversal operator. The projected
gap on the FS is shown in Fig.\ref{f}(a), revealing explicitly the $f$-wave symmetry.

We remark that during the FRG flow, there is a level crossing in the pp channel at the
energy scale indicated by the arrow in Fig.\ref{f}(b). The leading pairing symmetry is $p$-wave/$f$-wave above/below
the crossing point. This implies the possibility of $p$-wave pairing if the filling level is tuned further.

\begin{figure}
\includegraphics[width=8.5cm]{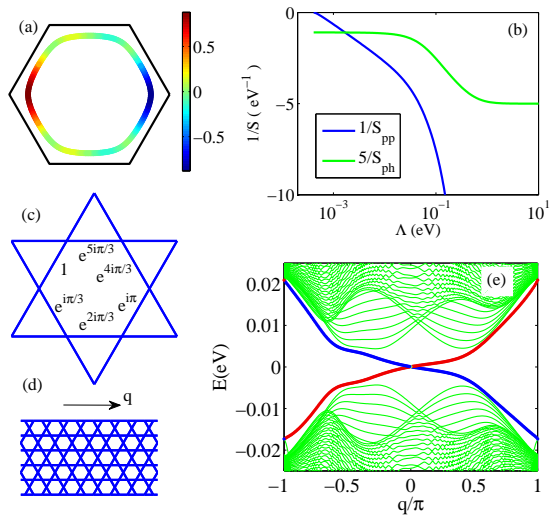}
\caption{ (Color online) Results for $\mu =0.02$eV. (a) The Fermi surface and the gap
function (color scale) of one of the degenerate pairing modes thereon. (b) FRG flow of (the inverse
of) the most negative singular values $S$ in the pp and ph channels. (c) The real-space pattern of the singlet part in the $p+ip'$ pairing.
(d) A strip of Kagome lattice periodic/open along the horizontal/vertical direction. Here $q$ denotes the conserved momentum. (e) The low
energy spectrum for (d) as a function of $q$. The red in-gap states are chiral modes on one edge, and the blue ones are on the other edge.}\label{pip}
\end{figure}

{\em $p+ip'$-wave superconductivity}: We consider further electron doping at $\mu = 0.02$eV. Fig.\ref{pip}(a)
and (b) show the corresponding FS and the FRG flow. As in the previous case, the ph
channel dominates at higher energy scales but saturates at lower scales, and the pp channel
is triggered attractive by the ph channel, and in turn diverges eventually. We find that there
are two degenerate paring modes in the pp channel. Resolving the matrix pairing function as before, we find the
dominant nonzero elements
\eqa
  g^{\al \bt}_{\v k} &=& - g^{\bt \al}_{\v k} = -i\sqrt{3} \Del_s \sin \frac{k_+}{2} , \nn
  g^{\al \ga}_{\v k} &=& - g^{\ga \al}_{\v k} = -i\sqrt{3} \Del_s \sin \frac{k_-}{2} , \nn
  d^{\al \al}_{\v k} &=& 2 \Del_t ( \sin k_+ - \sin k_- ) , \nn
  d^{\bt \bt}_{\v k} &=& \Del_t ( \sin k_x + \sin k_+ ),\nn
  d^{\ga \ga}_{\v k} &=& - \Del_t ( \sin k_x + \sin k_- ),
\eea
for the first mode $\phi_1(\v k)$, and
\eqa
  g^{\al \bt}_{\v k} &=& - g^{\bt \al}_{\v k} = -i \Del_s\sin \frac{k_+}{2} , \nn
  g^{\al \ga}_{\v k} &=& - g^{\ga \al}_{\v k} = i \Del_s \sin \frac{k_-}{2} , \nn
  g^{\bt \ga}_{\v k} &=& - g^{\ga \bt}_{\v k} = -2i \Del_s \sin \frac{k_x}{2} , \nn
  d^{\bt \bt}_{\v k} &=& \sqrt{3} \Del_t ( \sin k_x + \sin k_+ ),\nn
  d^{\ga \ga}_{\v k} &=& \sqrt{3} \Del_t ( \sin k_x + \sin k_- )
\eea
for the second mode $\phi_2(\v k)$. We find that $\Del_s : \Del_t \simeq 5 : 2$. So there is again a comparable mixture of
singlets and triplets.  We find $\phi_1(\v k)$ and $\phi_2(\v k)$ transform as $p_x$- and $p_y$-wave, respectively. As an
example, we project $\phi_1(\v k)$ on the FS in Fig.\ref{pip}(a) (color scale), where the $p_x$-wave symmetry is apparent.
The fact that the singlet part can transform as $p$-wave is again a consequence of the sublattice structure.

Since the two pairing modes are degenerate, additional analysis, such as the mean-field
theory, is needed to fix the gap function in the ordered state.
Given the explicit pairing functions from SMFRG, we are able to perform mean field
calculations by the following effective hamiltonian, in the momentum space,
\eqa H_{MF}=H_0 + \sum_{\v k,\nu} [\psi^\dag_\v k \Del_\nu \phi_\nu(\v k)(\psi^\dag_{-\v k})^T+{\rm h.c.}],\label{mft}\eea
subject to the self-consistency,
\eqa \Del_\nu = -g \<(\psi_{-\v k})^T\phi_\nu^\dag (\v k)\psi_\v k\>. \eea
Here $H_0$ is the free part of $H$, $\Del_\nu$ is the order parameter for $\phi_\nu(\v k)$, and
$g$ is an effective coupling constant fixing the transition temperature at the FRG divergence scale.
Notice that $\Del_\nu$ acts as the global amplitude for $\phi_\nu(\v k)$, so the
coefficients $\Del_{s,t}$ within $\phi_\nu(\v k)$ are fixed subject to the
given ratio from FRG. Our mean-field calculation reveals that in the ordered state,
the chiral combination $\phi_1(\v k) \pm i \phi_2(\v k)$ is always favorable.
The corresponding real-space pattern for the singlet part of the pairing function
is shown in Fig.\ref{pip}(c). This could have been anticipated since both $\phi_1(\v k)$
and $\phi_2(\v k)$ have nodes on the FS, a nature way to gain energy is to form a
$p\pm ip'$-wave pairing, which gaps out the entire FS.

The chiral $p+ip'$-pairing breaks time-reversal symmetry, and is topologically nontrivial.
The topology is classified by the Chern number, \cite{Schnyder}
\eq Z = \frac{1}{2\pi} \sum_{n} \int d^2 \v k (\p_{k_x} A^n_y - \p_{k_y} A^n_x)f(E_{n,\v k}), \ee
where $n$ is a band index, $A^n_i = -i \langle n,\v k | \p _{k_i} | n,\v k \rangle$ is the Berry connection, with
$ | n, \v k \rangle$ an eigenstate of $H_{MF}$ with energy $E_{n,\v k}$, and finally $f$ is the
Fermi function at zero temperature. We find $Z= 1$, similarly to the case for the $p+ip'$ pairing in Sr$_2$RuO$_4$. \cite{sro}
To verify the topology of our superconducting state,
we calculate the energy spectrum in a strip as shown in Fig.\ref{pip}(d).
The energy spectrum as a function of the horizontal momentum $q$ is shown in Fig\ref{pip}(e). Within the bulk energy gap (which is
artificially enlarged for a better view), there appear a branch of chiral states along each edge, in exact
correspondence to $Z=1$. Importantly, $H_{MF}$ can be embedded, as we did here, in a Nambu space without redundant degrees of
freedom. In this case each eigenstate describes a canonical fermion mode. Therefore the chiral edge states
are best termed as one-dimensional Weyl (rather than Majorana) fermion modes.\\

\begin{figure}
\includegraphics[width=7.5cm]{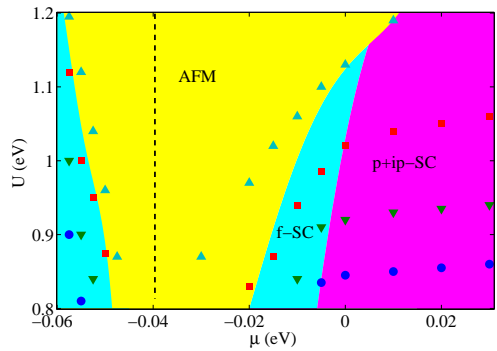}
\caption{ (Color online) A schematic phase diagram of $sd^2$-graphene in $(U, \mu)$ space.
 The dashed line highlights the van Hove level $\mu=-0.04$eV. The transition temperatures at
 some specific points of the phase diagram are $10$meV (upper triangles), $1$meV (squares), $0.1$meV (lower triangles),
 and $0.01$meV (cycles). } \label{pd}
\end{figure}

{\em The phase diagram}: Apart from the typical results discussed above,
we have performed systematic calculations for other values of $U$ and other
filling levels. The results are summarized as a schematic phase diagram in
Fig.\ref{pd}. With increasing deviation from the van Hove filling, the
system develops  intra-unitcell AFM, $f$-wave and $p+ip$-wave SC. With increasing
$U$, the fan of AFM increases, while the transition temperature (i.e., the divergence scale
in FRG) also increases in all phases.

We find the phase diagram is not changed qualitatively for $0.06$eV $\leq \la \leq 0.16$eV, except that
all the phase regimes are narrowed toward the van Hove filling if $\la$ is reduced.
The AFM survives down to $\la = 0$,\cite{wws2} but the transition temperature
for the SC states is too small to be of practical interest. On the other hand,
the system develops a CBO at large $U$ for $t_2 = t_3 = \la=0$ due to the
perfect nesting and matrix element effect.\cite{wws2} This state does not appear here
because the FS nesting is further degraded by $t_2$ and $t_3$.

\section{Summary and perspective} \label{SP}

We studied the electronic instabilities in electron-doped $sd^2$-graphene. 
With increasing deviation from the van Hove filling, the system develops $120^o$-type non-collinear intra-unitcell
AFM,  $f$-wave and $p+ip'$-wave SC states. In both SC states, singlets and triplets mix with comparable amplitudes due to SOC.
The chiral $p+ip'$-wave SC is characterized by a Chern number $Z =1$, supporting a branch of Weyl fermion modes along each edge.
Such modes can be probed by scanning tunneling microscopy, and would manifest itself as quantized thermal Hall conductivity.

\acknowledgments{The project was supported by NSFC (under grant Nos.11574134 and 11504085), 
the China Postdoctoral Science Foundation (under Grant No.2014M561616) and 
the Fundamental Research Funds for the Central Universities (under grant No. 2014B14314).}

\end{document}